\documentclass[runningheads]{llncs}
\usepackage[T1]{fontenc}
\usepackage[hidelinks, colorlinks=true, linkcolor=blue,citecolor = blue ,breaklinks = true]{hyperref}
\usepackage{graphicx,verbatim}
\usepackage{tikz}
\usepackage{diagbox}
\usepackage{amsmath}

\usepackage{amssymb}
\usepackage{longtable}
\usepackage{makecell}
\usepackage{latexsym}
\usepackage{enumitem}
\usepackage{pifont}
\usepackage{color}
\usepackage{titletoc}
\usepackage{colortbl}
\usepackage{wrapfig}  
\usepackage{lipsum} 
\usepackage{multirow}
\usepackage{listings}
\usepackage{array}
\usepackage{booktabs} 
\usepackage{caption}  

\usepackage{multicol}
\usepackage{bbm}
\usepackage{marvosym}
\usepackage{inconsolata}
\usepackage{ulem}
\usepackage{amsfonts}
\usepackage{algorithm}
\usepackage{algpseudocode}
\usepackage{float}

\definecolor{myGreen}{RGB}{34, 139, 34}
\definecolor{myRed}{HTML}{FF6347}

\newcommand{\cmark}{\textcolor{myGreen}{\ding{51}}}
\newcommand{\xmark}{\textcolor{myRed}{\ding{55}}}
\newcommand{\ours}{\textsc{MSWAL}} 
\definecolor{darkgreen}{RGB}{0,100,0}

\begin{document}
\title{MSWAL: 3D Multi-class Segmentation of Whole Abdominal Lesions Dataset}
\titlerunning{\textbf{MSWAL}}
\author{First Author\inst{1}\orcidID{0000-1111-2222-3333} \and
Second Author\inst{2,3}\orcidID{1111-2222-3333-4444} \and
Third Author\inst{3}\orcidID{2222--3333-4444-5555}}
\authorrunning{Zhaodong Wu et al.}
%

\institute{
Shanghai AI Lab, Shanghai, China
\and
University of Liverpool, Liverpool, UK
\and
Columbia university, New York, USA
\and
Monash University, Melbourne, Australia
\and
Xi'an Jiaotong-Liverpool University, Suzhou, China
\and
MBZUAI, Abu Dhabi, UAE
\and
People's Hospital of Gong'an County, Jingzhou City, Hubei Province
\\
Email: (\email{zhaodong@liverpool.ac.uk})
}

\author{Zhaodong Wu\inst{1,2}$^*$
\and
Qiaochu Zhao\inst{3}$^*$
\and
Ming Hu\inst{1,4}$^*$
\and
Yulong Li\inst{1,2,4}
\and\\
Haochen Xue\inst{1,2}
\and
Kang Dang\inst{5}
\and
Zhengyong Jiang\inst{5}
\and
Angelos Stefanidis\inst{5}
\and\\
Qiufeng Wang\inst{5}
\and
Imran Razzak\inst{6}
\and
Zongyuan Ge\inst{4}
\and
Junjun He\inst{1}
\and
Yu Qiao\inst{1}
\and\\
Zhong Zheng\inst{7}
\and
Feilong Tang\inst{1,4,6}
\and
Jionglong Su\inst{5}$^{(\textrm{\Letter})}$
}
    
\maketitle              

\def\thefootnote{$^{*}$}\footnotetext[1]{Equal contribution. }
\vspace{-8pt}
\begin{abstract}
With the significantly increasing incidence and prevalence of abdominal diseases, there is a need to embrace greater use of new innovations and technology for the diagnosis and treatment of patients. Although deep-learning methods have notably been developed to assist radiologists in diagnosing abdominal diseases, existing models have the restricted ability to segment common lesions in the abdomen due to missing annotations for typical abdominal pathologies in their training datasets. To address the limitation, we introduce \textbf{\ours}, the first 3D \textbf{M}ulti-class \textbf{S}egmentation of the \textbf{W}hole \textbf{A}bdominal \textbf{L}esions dataset, which broadens the coverage of various common lesion types, such as gallstones, kidney stones, liver tumors, kidney tumors, pancreatic cancer, liver cysts, and kidney cysts. With CT scans collected from 694 patients (191,417 slices) of different genders across various scanning phases, {\ours} demonstrates strong robustness and generalizability. The transfer learning experiment from {\ours} to two public datasets, LiTS and KiTS, effectively demonstrates consistent improvements, with Dice Similarity Coefficient (DSC) increase of 3.00\% for liver tumors and 0.89\% for kidney tumors, demonstrating that the comprehensive annotations and diverse lesion types in {\ours} facilitate effective learning across different domains and data distributions. Furthermore, we propose \textbf{Inception nnU-Net}, a novel segmentation framework that effectively integrates an Inception module with the nnU-Net architecture to extract information from different receptive fields, achieving significant enhancement in both voxel-level DSC and region-level F1 compared to the cutting-edge public algorithms on \ours. Our dataset will be released after being accepted, and the code is publicly released at  \url{https://github.com/tiuxuxsh76075/MSWAL-}.
\keywords{Dataset  \and Segmentation \and Abdominal Diseases.}

\end{abstract}
%
%
%
\newpage
\section{Introduction}
\label{sec:intro}
\vspace{-0.8em}
\sloppy 
Deep learning-based segmentation methods for abdominal lesions have significantly assisted radiologists in disease diagnosis~\cite{tang2024hunting,tang2024neighbor,zhao2024sfc,xu2024toward,xu2024polyp}, mitigating the rising demand for patient care and the limitations of available resources~\cite{schockel2020developments,winder2021we}. However, the performance of these methods is highly dependent on the scale and annotation quality of the datasets, where most datasets~\cite{bilic2023liver,heller2020international,antonelli2022medical} are annotated for lesions in a single organ, potentially leading to missed diagnoses by doctors. Specifically, the annotations in CT scan datasets such as LiTS~\cite{bilic2023liver}, KiTS~\cite{heller2020international}, and MSD task07~\cite{antonelli2022medical} focus on the lesions within individual organs, liver, kidney, and pancreas respectively. This single-organ lesions annotation paradigm may lead to diagnostic oversight, as it fails to account for potential lesions in other regions, thereby increasing the risk of missed diagnoses in clinical practice. 

Building on the single-organ annotation datasets, ULS'23~\cite{de2024uls23} and Flare'23~\cite{ma2024automatic} integrate a large number of public datasets and categorize all lesions into one type, pan-cancer, supporting a base to create algorithms to diagnose lesions in whole abdomen. Nonetheless, due to limitations in re-annotation during the integration of multiple datasets, many lesion labels are omitted in ULS'23~\cite{de2024uls23} and Flare'23~\cite{ma2024automatic}. For instance, ULS'23~\cite{de2024uls23} simply integrates KiTS~\cite{heller2020international} and LiTS~\cite{bilic2023liver}, where kidney tumors are not annotated in LiTS~\cite{bilic2023liver} and liver tumors are not annotated in KiTS~\cite{heller2020international}. Such annotation inconsistencies inevitably degrade model performance by introducing label noise in the training process. Moreover, the current approach classifies all lesions simply as pan-cancer, forcing doctors to spend extra time identifying the exact type of each lesion. Furthermore, only tumors are annotated in the abdomen, with other significant conditions, such as cysts and stones, are neglected. This oversight can result in missed diagnoses, affecting the accuracy, timeliness, and effectiveness of clinical decision-making.

Based on these limitations, we propose \textbf{\ours}, a 3D multi-class segmentation of whole abdominal lesions dataset, also the world's first large-scale common abdominal lesion segmentation dataset with accurate segmentation of different lesion types and fully annotated without missing labels. There are three advantages of {\ours} over the public datasets: \textit{(I)} \textbf{Common abdominal lesions}: {\ours} covers seven types of lesions, including gallstones, kidney stones, liver tumors, kidney tumors, pancreatic cancer, liver cysts, and kidney cysts. \textit{(II)} \textbf{Specific lesion types}: We annotate the lesions into seven types instead of labeled as pan-cancer. This fine-grained categorization enables deep learning models to generate more specific diagnostic suggestions, reducing radiologists' time in disease identification. \textit{(III)} \textbf{Full labels}: {\ours} is fully annotated, which reduces the impact of noise on deep learning methods and enhances their performance. Furthermore, we introduce \textbf{Inception nnU-Net} based on this dataset using Inception module~\cite{szegedy2015going}, consisting of parallel convolution layers with varying kernel sizes, enabling comprehensive feature extraction across varying lesion volumes. Leveraging this advantage, Inception nnU-Net achieves SOTA performance on \ours, demonstrating its ability in various lesions segmentation. 


The main contributions are listed as follows: \textit{(I)} We propose {\ours }, the first 3D multi-class segmentation of the whole abdominal lesions dataset. It has the advantage that seven types of common abdominal lesions are annotated on {\ours} with full labels and lesion types are categorized into various classes. \textit{(II)} The transfer learning experiment result demonstrates the superior generalization capability of {\ours} across diverse clinical scenarios, particularly in lesion segmentation tasks with varying anatomical contexts. \textit{(III)} The novel model Inception nnU-Net fusing the advantages of nnU-Net architecture and Inception module \cite{szegedy2015going} to extract information on different types of lesions is proposed. Our comparative experiment shows that it achieves SOTA performance on \ours, which validates its effectiveness. \textit{(IV)} We establish a new benchmark on {\ours} to assess the performance of six cutting-edge methods.

\section{\ours}
\subsubsection{Dataset Statistical Analysis.}

\begin{figure}[!t]
    \centering
    \includegraphics[width=\textwidth]{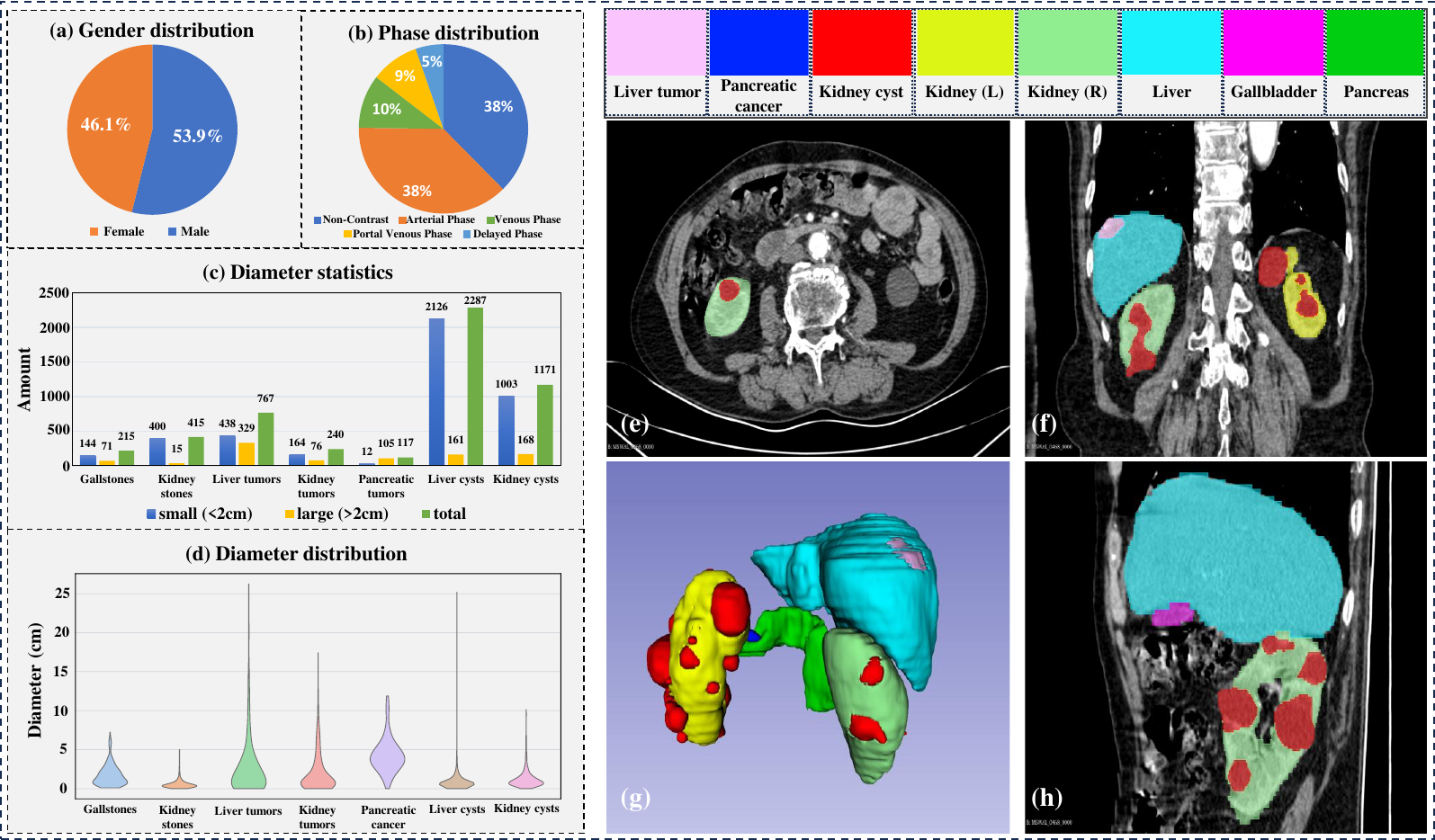}
    \caption{Data distributions and an example of \ours. (a) gender distribution (male and female); (b) phase distribution (non-contrast, arterial phase, venous phase, portal venous phase, and delayed phase); (c) diameter statistics. The Lesions are categorized into large and small based on a diameter threshold of 2 cm; (d) diameter distribution; (e)-(h) an example of \ours: (e) axial plane; (f) coronal plane; (g) 3D show; (h) sagittal plane.}
    \label{figure1: distribution and example}
    \vspace{-10pt}
\end{figure}

{\ours} contains 694 CT scans (191,417 slices) from 694 patients at one hospital. Each CT volume consists of 52 to 1,089 slices of $512 \times 512$ pixels, demonstrating the high resolution of {\ours}. The majority of the data are acquired by a 64-slice Computed Tomography scanner manufactured by GE Healthcare, with a small subset collected from other imaging devices. The dataset is randomly split into a training set of 484 volumes and a testing set of 210 volumes, maintaining an approximate 7:3 ratio. As Fig.~\ref{figure1: distribution and example} (a) shows, {\ours} exhibits a balanced gender distribution in our dataset, with 374 (53.9\%) male and 320 (46.1\%) female patients. Furthermore, {\ours} also has comprehensive coverage across multiple CT contrast phases, as illustrated in Fig.~\ref{figure1: distribution and example} (b), including non-contrast (261, 38\%), arterial phase (261, 38\%), venous phase (71, 10\%), portal venous phase (64, 9\%), and delayed phase (37, 5\%). Beyond demographic and imaging protocol diversity, {\ours} contains detailed annotations for seven clinically significant lesion types: gallstones (215 instances), kidney stones (415 instances), liver tumors (767 instances), kidney tumors (240 instances), pancreatic tumors (117 instances), liver cysts (2,287 instances), and kidney cysts (1,171 instances). Following the WHO measurement standard, lesions are categorized by size, with those exceeding 2 cm in diameter classified as large, as visualized in Fig.~\ref{figure1: distribution and example} (c). The corresponding size distribution across all lesion types is presented in Fig.~\ref{figure1: distribution and example} (d). Fig. \ref{figure1: distribution and example} (e-h) display three orthogonal views of an example and a 3D visualization from our dataset. Although we do not annotate the organs, their labels are inferred by nnU-Netv1 trained on WORD dataset \cite{luo2022word} to display our annotation of lesions more clearly. Furthermore, to highlight the advantages of \ours, comparisons between {\ours} and public datasets are shown in Table \ref{table1: comparison between public datasets and MSWAL}.

\begin{table}[!t]

\centering

\fontsize{8}{10}\selectfont
\caption{The comparison between public datasets and \ours.}
\resizebox{\linewidth}{!}{
\begin{tabular}{l c c c@{\hspace{8pt}} c@{\hspace{8pt}} c@{\hspace{8pt}} c@{\hspace{10pt}} c@{\hspace{10pt}} c @{\hspace{10pt}}c @{\hspace{10pt}}c @{\hspace{10pt}}c @{\hspace{10pt}}c @{\hspace{10pt}}c @{\hspace{10pt}}c}
\midrule
\hline
\textbf{dataset}&\textbf{CTs} & \textbf{AL} & \textbf{NSD} & \textbf{FL} & \textbf{SLT} & \textbf{GS} & \textbf{KS} & \textbf{LT} & \textbf{KT} & \textbf{PC} & \textbf{LC} & \textbf{KC}
\\
\hline
BTCV \cite{landman2015miccai} &50 &Null&  \cmark&  \cmark&  \xmark&  \xmark&  \xmark&  \xmark&  \xmark&  \xmark&  \xmark&  \xmark \\
CHAOS~\cite{kavur2021chaos}&20&Null&\cmark &\cmark &\xmark&  \xmark&  \xmark&  \xmark&  \xmark&  \xmark&  \xmark&  \xmark\\
CT-ORG~\cite{rister2020ct}&140&Null&\xmark&\cmark&\xmark&  \xmark&  \xmark&  \xmark&  \xmark&  \xmark&  \xmark&  \xmark\\
WORD \cite{luo2022word}&170&Null&  \cmark&  \cmark&  \xmark&  \xmark&  \xmark&  \xmark&  \xmark&  \xmark&  \xmark&  \xmark\\
RAOS \cite{luo2024rethinking}&413&Null&  \xmark&  \cmark&  \xmark&  \xmark&  \xmark&  \xmark&  \xmark&  \xmark&  \xmark&  \xmark
\\
AbdomenAtlas 1.1 \cite{li2024abdomenatlas} & 20,460 & Null & \xmark& \cmark& \xmark& \xmark& \xmark& \cmark& \cmark& \cmark& \xmark& \xmark\\
LiTS \cite{bilic2023liver} & 201 & 1,372\textbf{*} & \cmark & \cmark&  \cmark &  \xmark &  \xmark & \cmark &  \xmark &  \xmark &  \xmark &  \xmark \\
KiTS \cite{heller2020international} & 599 & 1,406\textbf{*} &  \cmark & \cmark &  \cmark &  \xmark& \xmark& \xmark& \cmark& \xmark& \xmark& \cmark\\
MSD task07 \cite{antonelli2022medical} & 420 &  423\textbf{*} &  \cmark & \cmark & \cmark &  \xmark& \xmark& \xmark& \xmark& \cmark& \xmark& \xmark\\
ULS'23 \cite{de2024uls23} & 6,994 & 1,618 & \xmark& \xmark& \xmark& \xmark& \xmark& \xmark& \xmark& \xmark& \xmark& \xmark\\
Flare'23 \cite{ma2024automatic} & 4500 & $\sim$ & \xmark& \xmark& \xmark& \xmark& \xmark& \xmark& \xmark& \xmark& \xmark& \xmark\\
\midrule
\textbf{{\ours} (Ours)}& 694 & 5,212 & \cmark& \cmark& \cmark& \cmark& \cmark& \cmark& \cmark& \cmark& \cmark& \cmark\\
\midrule
\hline
\end{tabular}
}
 \raggedright
 \scriptsize
  \textbf{AL}: \textbf{Annotated Lesions} refers to the number of lesions labeled by the authors, excluding those annotated before from other datasets. This distinguishes our work from datasets that primarily rely on external annotations. \textbf{NSD}: \textbf{New Source Data} means whether all original volumes are released at the first time. \textbf{FL}: \textbf{Full Label} means there is no missing annotation on the target regions. \textbf{SLT}: \textbf{Specific Lesion type} is given \cmark if diseases are categorized into specific types instead of pan-cancer. \textbf{GS}: Gallstones; \textbf{KS}: Kidney Stones; \textbf{LT}: Liver Tumors; \textbf{KT}: Kidney Tumors; \textbf{PC}: Pancreatic Cancer; \textbf{LC}: Liver Cysts; \textbf{KC}: Kidney Cysts. \textbf{*} displays incomplete statistics because the paper does not give this information. $\boldsymbol{\sim}$ means the data can not be calculated. 
\label{table1: comparison between public datasets and MSWAL}
\vspace{-10pt}
\end{table}
\vspace{-5pt}
\subsubsection{Ground Truth Generation.}
During the process of dataset collection, all data are validated by the ethics committee under approval number XXX (anonymized requirements). To guarantee the high quality of \ours, all CT scans are annotated by an attending physician (with over 10 years of clinical experience) using 3D slicer's semi-automatic function~\cite{fedorov20123d}. Specifically, after annotating several slices in the axial, sagittal, and coronal views, the physician generates the initial 3D annotations, which are then carefully adjusted to modify the boundaries. Subsequently, another chief physician (with over 20 years of clinical experience) checks the annotations and further discusses the inconsistent ideas of lesion regions with the attending physician. It takes 11 months to label and review the \ours. According to our statistics, each volume averagely takes 1.1 hours to annotate and requires 0.3 hours to review and discuss. It is worth noticing that the doctors categorize lesions into seven types, the dataset has no missing labels, and all original volumes are released for the first time. 


\section{Inception nnU-Net}

\begin{figure}[t]
    \centering
    \includegraphics[width=\textwidth]{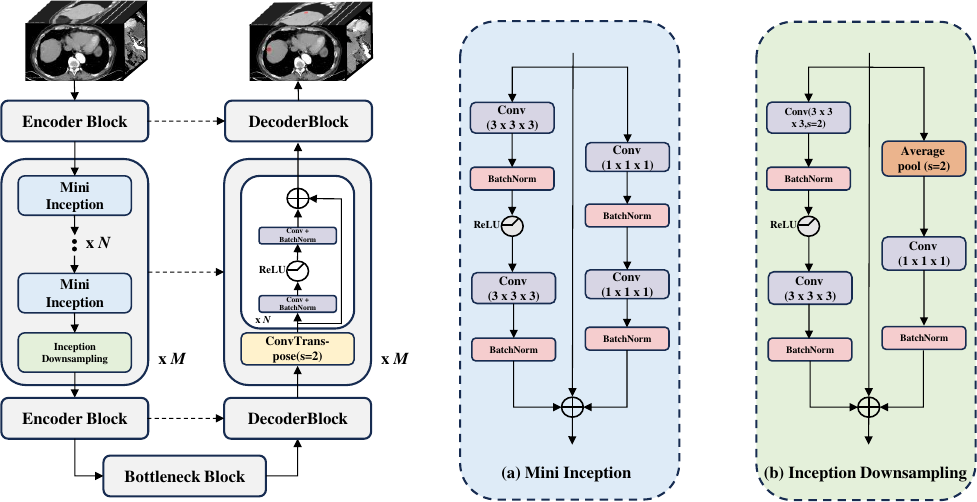}
    \caption{The left architecture is Inception nnU-Net, whose Bottleneck Block is Encoder Block apart from Inception Downsampling. (a) Mini Inception, a component of Inception nnU-Net (b) Inception Downsampling, another module of Inception nnU-Net. Both Mini Inception and Inception Downsampling have two branches, called branch left and branch right.}
    \label{figure2: model}
    \vspace{-18pt}
\end{figure}

\subsubsection{Mini Inception.}
The Inception module~\cite{szegedy2015going} employs convolutional layers with varying kernel sizes to extract multi-scale features from different receptive fields. Inspired by this design, we propose Mini Inception, a compact module that retains the core functionality of the original Inception module~\cite{szegedy2015going} while simplifying its structure. As shown in Fig. \ref{figure2: model} (a), Mini Inception incorporates a residual connection, represented as the central arrow, and two parallel branches: In the left branch, a $3 \times 3 \times 3$ convolutional layer is applied, followed by batch normalization and a ReLU activation layer. Subsequently, another $3 \times 3 \times 3$ convolutional layer is employed, accompanied by batch normalization. In the right branch, a $1 \times 1 \times 1$ convolutional layer is utilized to capture features from different receptive fields, followed by batch normalization. This is then succeeded by an additional $1 \times 1 \times 1$ convolutional layer and batch normalization. Compared to the original Inception module \cite{szegedy2015going}, Mini Inception reduces the number of branches to two, utilizing only two kernel sizes ($3 \times 3 \times 3$ and $1 \times 1 \times 1$) while maintaining the ability to extract multi-scale information effectively. 



\vspace{-20pt}
\subsubsection{Inception Downsampling.}
To capture multi-scale features during downsampling, we introduce Inception Downsampling, a novel module designed to extract information from various receptive fields. Like Mini Inception, the central arrow in Fig.~\ref{figure2: model} (b) represents the residual connection. The module consists of two parallel branches for feature extraction: In the left branch, a convolutional layer with a stride of 2 is employed for downsampling, followed by batch normalization. Subsequently, a ReLU activation, a $3 \times 3 \times 3$ convolutional layer, and batch normalization are applied. An average pooling layer with a stride of 2 is used in the right branch, followed by a $1 \times 1 \times 1$ convolutional layer and batch normalization. This design allows Inception Downsampling to effectively aggregate multi-scale information, making it suitable for segmenting lesions of varying sizes.

\vspace{-12pt}

\section{Experiments and Results}

\subsubsection{Implementation Details.}
All experiments are conducted using the RTX 4090. Our proposed model, Inception nnU-Net, uses nnU-Net's pre-processing with an initial learning rate of 0.001 and a total of 1,500 training epochs. In our training process, we employed a linear decay learning rate schedule to adjust the learning rate. In the evaluation stage, two metrics are used: voxel overlap-based metric Dice similarity coefficient (DSC) and region-level F1 with an Intersection over Union (IOU) threshold of 0.5. It is worth noting that region-level F1 is used because we observed that there are many region-level false positives and false negatives during inference. Compared to predicting the wrong boundaries, predicting the wrong regions will bring more misleading information for doctors. This is why the region-level F1 score, highly sensitive to region-level false negatives and false positives, is adopted as a key metric for model evaluation.

\begin{table}[t]
\vspace{-15pt}
\centering
\fontsize{8}{10}\selectfont
\caption{The comparative experiment between Inception nnU-Net and six SOTA methods in terms of DSC(\%) and F1(\%) with IOU (threshold = 0.5) on \ours. The average score means the mean value of Inception nnU-Net and six SOTA methods on MSWAL. We calculate the average score to evaluate the relative difficulty of various types of lesions segmentation, with lower average scores indicating harder diagnostic challenges for specific diseases. \textbf{Bold} and \underline{underline} denote the best and second-best results, respectively.}
\resizebox{\linewidth}{!}{
\begin{tabular}{l| c| c| c| c| c| c| c| c |c|c |c| c |c| c|c|c}
\hline
\hline
\multirow{3}{*}{{\diagbox{\makecell{lesion}}{method}} }
 & \multicolumn{2}{c|}{\multirow{2}{*}{\makecell{Inception\\ nnU-Net (Ours)}}}& \multicolumn{2}{c|}{\multirow{2}{*}{\makecell{nnU-Netv1\\~\cite{isensee2021nnu}}}} & \multicolumn{2}{c|}{\multirow{2}{*}{\makecell{nnU-Netv2\\ \cite{isensee2021nnu}}}} & \multicolumn{2}{c|}{\multirow{2}{*}{\makecell{nnU-Net\\ res~\cite{isensee2024nnu}}}}& \multicolumn{2}{c|}{\multirow{2}{*}{Mednext~\cite{roy2023mednext}}} & \multicolumn{2}{c|}{\multirow{2}{*}{nnFormer~\cite{zhou2023nnformer}}} & \multicolumn{2}{c|}{\multirow{2}{*}{\makecell{Swin\\ UNETR~\cite{hatamizadeh2021swin}}}}&\multicolumn{2}{c}{\multirow{2}{*}{\makecell{Average\\score}}}
\\
&\multicolumn{2}{c|}{}&\multicolumn{2}{c|}{}&\multicolumn{2}{c|}{}&\multicolumn{2}{c|}{}&\multicolumn{2}{c|}{}&\multicolumn{2}{c|}{}&\multicolumn{2}{c|}{}&\multicolumn{2}{c}{}
\\ \cline{2-17}
& DSC& F1 & DSC& F1 & DSC& F1 & DSC& F1 & DSC& F1 & DSC& F1 & DSC& F1  & DSC& F1 
\\
\hline
Gallstone&\underline{57.12}&52.46&45.54&44.36&33.60&38.26&55.95&57.81&60.27&64.61&42.96&38.23&36.48&36.81&47.41&47.50\\

Kidney stone &35.75&25.67&22.86&15.83&23.13&16.70&31.47&23.70&37.11&32.46&37.84&31.98&33.84&23.96&31.71&24.32\\

Liver tumor&$\textbf{58.53}_{\textcolor{darkgreen}{\uparrow 2.53}}$&$\textbf{23.05}_{\textcolor{darkgreen}{\uparrow 2.48}}$&29.42&10.23&40.74&14.22&56.00&20.57&45.10&17.50&42.51&18.15&27.52&11.37&42.83&16.44\\

Kidney tumor&44.08&\underline{27.69}&21.05&12.24&40.47&18.67&45.41&29.00&46.58&22.11&37.82&26.11&16.02&23.13&35.91&22.70\\

Pancrea cancer&\underline{48.77}&$\textbf{44.44}_{\textcolor{darkgreen}{\uparrow 3.54}}$&27.28&23.89&37.40&36.20&54.64&40.90&43.68&37.89&42.20&39.04&19.43&20.98&39.05&34.76\\

Liver cyst &$\textbf{58.47}_{\textcolor{darkgreen}{\uparrow 2.23}}$&$\textbf{45.51}_{\textcolor{darkgreen}{\uparrow 1.74}}$&38.59&37.67&36.56&40.45&56.24&43.77&41.59&41.69&46.85&40.80&37.16&33.89&45.06&40.54\\

Kidney Cyst &$\textbf{47.87}_{\textcolor{darkgreen}{\uparrow 0.19}}$&$\textbf{58.14}_{\textcolor{darkgreen}{\uparrow 1.71}}$&41.95&46.47&40.90&50.39&47.68&56.43&45.23&51.33&30.53&50.76&28.14&33.92&40.32&49.63\\
\hline
Avg. &$\textbf{50.09}_{\textcolor{darkgreen}{\uparrow 0.46}}$&$\textbf{39.56}_{\textcolor{darkgreen}{\uparrow 0.68}}$&32.38&27.24&36.11&30.69&49.63&38.88&45.65&38.22&40.10&35.01&28.37&26.29&40.32&33.69\\
\hline
\hline

\end{tabular}}
\vspace{-10pt}
\label{table:DSC comparison}

\end{table}

\subsubsection{Comparative Experiment.}
\begin{figure}[!t]
    \centering
    \includegraphics[width=\textwidth]{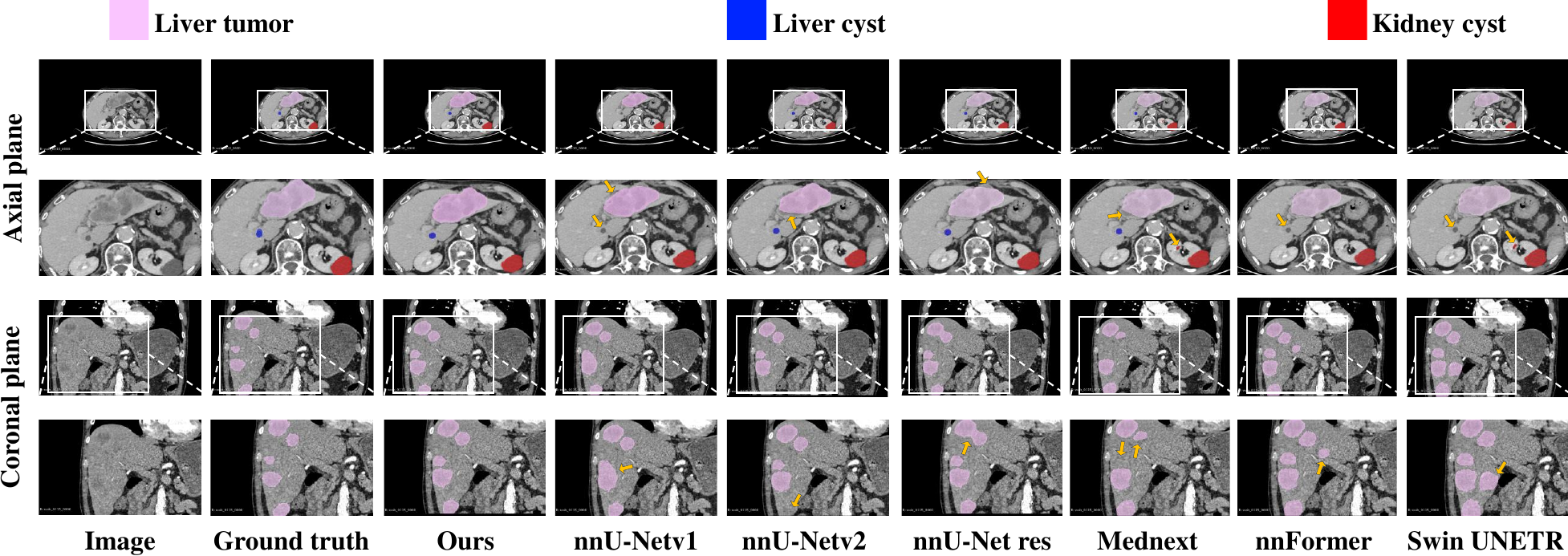}
    \caption{Visualization of the segmentation results from different methods. The yellow arrows point to the shortcomings of the six cutting-edge models in terms of edge segmentation, or to the regions of false positives and false negatives.}
    \label{figure: display an example}
    \vspace{-10pt}
\end{figure}

We investigate Inception nnU-Net and six public SOTA 3D medical image segmentation methods on \ours, including nnU-Netv1~\cite{isensee2021nnu}, nnU-Netv2~\cite{isensee2021nnu}, nnU-Net res~\cite{isensee2024nnu}, Mednext~\cite{roy2023mednext}, nnFormer~\cite{zhou2023nnformer}, and Swin UNETR~\cite{hatamizadeh2021swin} as shown in Table. \ref{table:DSC comparison}. The comparative experiment clearly shows Inception nnU-Net's advantages over the existing models, as illustrated by the example in Fig. \ref{figure: display an example}. With existing methods exhibiting significant limitations in accurately predicting blurry boundaries and often generating false negative and false positive regions, our proposed method, Inception nnU-Net, demonstrates superior performance in region and boundary prediction. Notably, our analysis based on average scores reveals that segmenting kidney stones and kidney tumors is the most challenging task among the seven types of lesions studied. This difficulty persists even though pancreatic cancer, which is clinically considered the most complex condition~\cite{siegel2023cancer,cao2023large}, achieves or approaches the average DSC and F1 scores. The relatively better performance in pancreatic cancer segmentation can be attributed to the simpler annotation scheme, as only one type of lesion is annotated in the pancreas. In contrast, the kidney presents a more complex scenario with three distinct lesion types: kidney stones, tumors, and cysts. Crucially, this complexity is further compounded by a significant class imbalance in the dataset, where kidney cysts are overrepresented (1,171 instances) compared to kidney stones (415 instances) and kidney tumors (240 instances). The disproportionate distribution of lesion types and the mutual influence of multiple lesions within a single organ present a particularly significant challenge in our study (\ours). This interplay of factors underscores the complexity of multi-lesion segmentation tasks in medical imaging analysis. Therefore, based on this, we suggest that future researchers carefully explore the issues of mutual interference among lesions and the long-tail problem in the MSWAL task.
\vspace{-8pt}
\subsubsection{Ablation Experiment.}
We evaluate the impact of different modules in Inception nnU-Net on medical segmentation performance in {\ours} as Table~\ref{table:Ablation experiments} shows. The baseline model, illustrated in variant I, removes two key modules: branch left in Inception Downsampling and branch right in Mini Inception, and the residual connection in Mini Inception. This results in a significant DSC drop by 1.07\%, highlighting their importance. In variant II, removing only the branch left in Inception Downsampling reduces the DSC to 49.46\%. This result shows that compared to other components, this one is the most critical, as it significantly increases the DSC value by 0.63\%. Variant III involves removing the branch right in Mini Inception, which decreases the DSC to 48.83\%. The 0.26\% improvement in DSC underscores the importance of this component. Variant IV shows that removing the residual connection in Mini Inception causes a minor DSC decrease by 0.07\%. Finally, in variant V, restoring all components achieves the highest DSC score, 50.09\%. This validates the effectiveness of the full Inception nnU-Net model, particularly underscoring the critical roles of Inception Downsampling and Mini Inception modules in enhancing segmentation accuracy.


\begin{table}[t]
\vspace{-10pt}
\centering
\fontsize{8}{10}\selectfont
\caption{Ablation experiment of Inception nnU-Net in terms of DSC(\%) and F1(\%) with IOU (threshold = 0.5) on \ours. Inception Downsampling: Branch left in Inception Downsampling. Mini Inception: Branch right in Mini Inception. Residual connection: Residual connection in Mini Inception.}
\begin{tabular}{c|c|c|c|c|c}
\hline
\hline
\multirow{2}{*}{\makecell{\textbf{Variants}}}&\multirow{2}{*}{\makecell{Inception\\ Downsampling}}&\multirow{2}{*}{\makecell{Mini\\Inception}}& \multirow{2}{*}{\makecell{Residual\\connection}}&\multirow{2}{*}{\makecell{DSC}}&\multirow{2}{*}{\makecell{F1}}\\
&&&&&
\\
\hline
I&\xmark&\xmark&\xmark&49.02&38.39\\ \hline
II&\xmark&\cmark&\cmark&49.46&38.78\\ \hline
III&\cmark&\xmark&\cmark&49.83&39.23\\ \hline
IV&\cmark&\cmark&\xmark&50.02&39.42\\ \hline
V&\cmark&\cmark&\cmark&50.09&39.56\\ 
\hline
\hline

\end{tabular}
\label{table:Ablation experiments}
\vspace{-10pt}
\end{table}

\vspace{-10pt}
\subsubsection{Transfer Learning Experiment.} 
Models trained on extensive and high-quality datasets are expected to have enhanced transfer learning capabilities, improving generalization across relevant sub-domains. We conduct the transfer learning experiment to validate the efficiency of {\ours} for algorithms in transfer learning scenarios. We choose the public SOTA method, nnU-Net res~\cite{isensee2024nnu} as the baseline method with two training configurations: \textit{(I)}: the model is trained on LiTS~\cite{bilic2023liver} and KiTS~\cite{heller2020international} without transfer learning. \textit{(II)}: the model is firstly pre-trained on MSWAL then fine-tune on LiTS~\cite{bilic2023liver} and KiTS~\cite{heller2020international}. Since the official testing sets of LiTS and KiTS are not released, we split the training sets into new training sets and test sets in an 8:2 ratio. As shown in Table. \ref{table: Transfer learning}, we find that although the organs segmentation improvement is not obvious, i,e., DSC of the liver in LiTS increases from 96.79\% to 96.82\% and kidney in KiTS increases from 95.42\% to 95.49\%, the lesions segmentation improvement is significant. In LiTS, the DSC of liver tumors increases from 71.40\% to 74.40\%. In KiTS, the DSC of kidney tumors increases from 88.50\% to 89.39\%, and kidney cyst increases from 43.57\% to 44.04\%. The great improvement demonstrates {\ours}'s robustness and generalization, especially for the abdominal lesion segmentation.
\begin{table}[t]
\vspace{0pt}
\centering
\fontsize{8}{10}\selectfont
\caption{Transfer learning from {\ours} to LiTS and KiTS using nnU-Net res.}
\begin{tabular}{l|c|c|c|c|c}
\hline
\hline
Dataset&\multicolumn{2}{c|}{{\ours} $\rightarrow$ LiTS}&\multicolumn{3}{c}{{\ours} $\rightarrow$ KiTS}
\\
\hline
Task&Liver & Liver tumor & Kidney & Kidney tumor & Kidney cyst
\\
DSC&96.79 $\rightarrow$ 96.82 & 71.40 $\rightarrow$ 74.40 & 95.42 $\rightarrow$ 95.49 & 88.50 $\rightarrow$ 89.39 & 43.57 $\rightarrow$ 44.04 
\\
Improve&\textcolor{darkgreen}{${\uparrow}$0.03} &\textcolor{darkgreen}{${\uparrow}$3.00} &\textcolor{darkgreen}{${\uparrow}$0.07 }&\textcolor{darkgreen}{${\uparrow}$0.89 }&\textcolor{darkgreen}{${\uparrow}$0.47} \\
\hline
\hline

\end{tabular}
\label{table: Transfer learning}
\vspace{-15pt}
\end{table}

\vspace{-0.5em}
\section{Conclusion}
\vspace{-0.5em}
In this paper, we introduce {\ours}, a large-scale dataset comprising 694 patients (191,417 slices) for segmenting seven types of abdominal lesions. To highlight its robustness and generalization, we explore the domain gap between {\ours} and two public datasets (LiTS and KiTS). Furthermore, we propose the novel Inception nnU-Net framework, designed to capture information across varying receptive fields. The comparative experiment demonstrates the effectiveness of Inception nnU-Net, with improvements of 0.46\% in DSC and 0.68\% in F1-score, showcasing its superior region-level segmentation capabilities on {\ours}. However, the lack of integration with whole abdominal lesion report generation tasks is a limitation of {\ours}, which we aim to address in future work to enhance the dataset's utility and comprehensiveness.


%
%
%
%
\bibliographystyle{splncs04}  
\bibliography{references}   
\end{document}